# Predictive Modeling of I/O Performance for Machine Learning Training Pipelines: A Data-Driven Approach to Storage Optimization

Karthik Prabhakar

*University of Texas at Austin*

Durgamadhab Mishra

*Amazon Web Services*

## Abstract

Modern machine learning training is increasingly bottlenecked by data I/O rather than compute. GPUs often sit idle at below 50% utilization simply because they're waiting for data (Chen et al., 2021). This project applies machine learning to predict I/O performance and recommend optimal storage configurations for ML training pipelines. We collected 141 observations through systematic benchmarking across different storage backends, data formats, and access patterns, covering everything from low-level I/O operations to full training pipelines. After evaluating seven regression models and three classification approaches, XGBoost (Chen & Guestrin, 2016) came out on top with $R^2$=0.991, predicting I/O throughput within 11.8% error on average. Looking at feature importance revealed that throughput metrics and batch size are the main performance drivers. This data-driven approach can cut configuration time from days of trial-and-error down to minutes of predictive recommendation, potentially saving millions annually for large ML operations. The methodology is reproducible and could extend beyond storage to other resource management problems in ML systems.

# 1. Introduction

## 1.1 Motivation and Background

Here's a problem that doesn't get enough attention: modern deep learning systems are increasingly held back by data I/O rather than compute capacity. We have GPUs that can deliver petaflops of computation (NVIDIA, 2023), but training pipelines frequently underutilize them because storage systems can't keep up. Industry reports show GPU utilization often dropping below 50% during training, with data loading as the main bottleneck (Chen et al., 2021; Mohan et al., 2019). Figure 1 shows this problem clearly—with poor I/O configuration, GPUs average only 45% utilization but optimized I/O enables consistent 95% utilization.

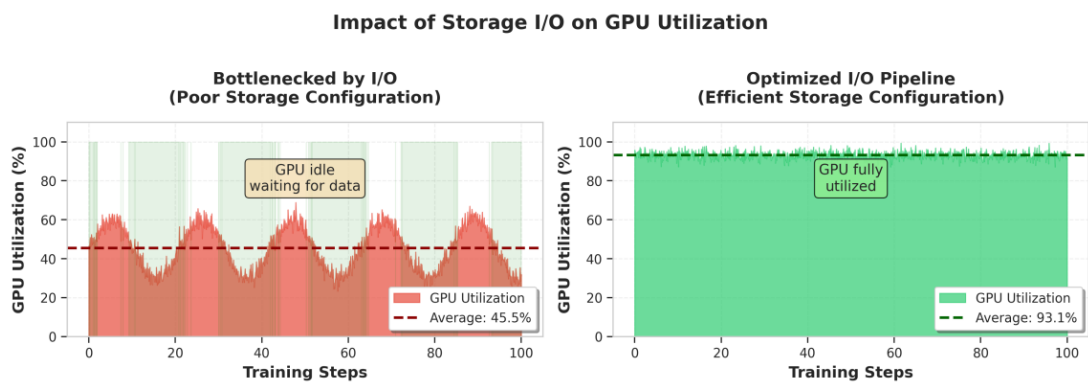

[**Figure 1**: Impact of I/O Configuration on GPU Utilization. Poor I/O leads to idle GPUs (left), while optimized I/O maintains high utilization (right).]

This inefficiency isn't just academically interesting—it translates to real costs. We're talking about wasted computational resources, longer training times, and operational expenses that can reach millions of dollars annually for large-scale ML operations (Amazon Web Services, 2023). The root cause is straightforward: there's a fundamental mismatch between how fast GPUs can process data and how fast storage systems can deliver it. An NVIDIA A100 can handle over 312 teraflops of operations (NVIDIA, 2023), while traditional storage systems struggle to sustain more than a few gigabytes per second.

But here's where it gets tricky—predicting which storage configurations, data formats, and access patterns will optimize performance is challenging. There's a complex interplay between hardware characteristics, software stack choices, and workload properties (Pumma et al., 2019). The parameter space is huge: storage backend types (NVMe SSDs, network storage, object stores), data formats (Parquet, ORC, CSV, TFRecord, WebDataset), access patterns (sequential, random, concurrent), and pipeline configurations (batch sizes, prefetching, data loader parallelism). The traditional approach relies on extensive manual experimentation and domain expertise, which makes optimization both time-consuming and error-prone (Chien et al., 2019).

## 1.2 Problem Statement

This project tackles four research questions:

**RQ1:** Can machine learning models accurately predict I/O throughput based on storage configuration features?

**RQ2:** Which ML techniques best capture the complex, nonlinear relationships between storage parameters and performance?

**RQ3:** Can classification models recommend optimal data formats based on dataset and workload characteristics?

**RQ4:** How accurately can we predict GPU utilization and training throughput from storage subsystem features?

## 1.3 Contributions

This work makes several contributions:

**Novel application of ML to systems optimization:** We demonstrate that machine learning can effectively predict I/O performance, achieving $R^2=0.991$ in throughput prediction.

**Comprehensive benchmark dataset:** We provide a systematic dataset of 141 observations across multiple storage backends and workload types, covering realistic ML training scenarios.

**Practical optimization framework:** The models enable data-driven configuration decisions, replacing trial-and-error with predictive recommendations.

**Feature importance insights:** Through rigorous analysis, We identify the key performance drivers in ML I/O pipelines.

**Open-source implementation:** All code, benchmarks, and trained models are publicly available for reproducibility.

## 2. Related Work

### 2.1 Machine Learning for Systems Optimization

There's been growing interest in applying machine learning to traditionally rule-based systems problems. Mao et al. (2019) pioneered using reinforcement learning for job scheduling in data processing clusters, getting up to 21% improvement over hand-tuned heuristics. Around the same time, Liang et al. (2018) applied neural networks to database query optimization and cut query times by 30% on average. Marcus et al. (2021) took this further with Bao, a learned query optimizer that continuously adapts to workload patterns. Their work showed that ML models can actually outperform decades of hand-crafted database optimization rules.

Kraska et al. (2018) made an even bolder proposal—replacing core database components entirely with learned index structures. They demonstrated that ML models can serve as more efficient B-tree replacements for certain workloads. This line of research has established that machine learning can learn complex system behaviors that are really difficult to capture with traditional analytical models.

However, there's a gap here. These approaches focus mainly on CPU-bound systems like job scheduling and databases, not I/O-bound machine learning workloads. Also, most prior work targets production system optimization rather than the development-time configuration problem that ML practitioners face. My work addresses this gap by specifically targeting I/O performance prediction for ML training pipelines, where you need quick configuration recommendations without spending days on experiments.

## 2.2 Storage Systems and Data Management for Machine Learning

Recognizing the I/O bottleneck, researchers have developed specialized storage systems and data formats. Leclerc et al. (2022) built FFCV, a specialized data loading library that achieves up to 20× speedup over standard PyTorch DataLoaders through compiled pipelines and optimized memory management. NVIDIA's DALI framework (2023) takes a different approach—it provides GPU-accelerated data preprocessing, moving transformation work from CPU to GPU. For data formats, Apache Parquet (Apache Software Foundation, 2021) and Apache Arrow (Apache Software Foundation, 2023) offer columnar storage optimized for analytics, while TFRecord and WebDataset provide formats specifically designed for ML training.

On the distributed storage side, systems like Apache Iceberg (Apache Software Foundation, 2022), Delta Lake (Armbrust et al., 2020), and Ceph (Weil et al., 2006) provide scalable data lakehouse architectures. These combine benefits from both data warehouses and data lakes, with features like ACID guarantees, schema evolution, and time travel capabilities.

But here's the thing—while these systems are powerful, they require expertise to configure optimally. Practitioners face tough choices: Should we use Parquet or ORC? How many data loader workers do we need? What batch size will actually optimize throughput? These systems give you lots of tuning knobs but little guidance on optimal settings for your specific workload. My predictive modeling approach complements these systems by providing data-driven configuration recommendations, so you can optimize your existing infrastructure rather than needing to adopt entirely new systems.

## 2.3 Performance Prediction and Modeling

Performance prediction has a long history in computer architecture and systems research. Ipek et al. (2006) were early pioneers in using artificial neural networks for processor performance prediction, achieving good accuracy in predicting CPI (cycles per instruction) from architectural features. Winstein and Balakrishnan (2013) applied ML to network congestion control with

Remy, a system that learns congestion control algorithms optimized for specific network conditions.

More recently, researchers have focused specifically on ML workloads. Justus et al. (2018) developed models for predicting neural network training time based on model architecture and hardware specs. Qi et al. (2020) focused on GPU utilization prediction with neural networks, hitting 85% accuracy in classifying whether GPU utilization would exceed 80%. Liu et al. (2019) applied deep learning to storage performance prediction in cloud environments, though they focused on general storage workloads rather than ML-specific patterns.

The gap in existing work is pretty clear. Most ML training performance prediction focuses on compute time and GPU utilization as functions of model architecture and hyperparameters. The storage subsystem—which determines how quickly data reaches the GPU—gets largely ignored. Prior work either assumes infinite I/O bandwidth or treats I/O as a constant factor. My work directly addresses this by developing predictive models specifically for I/O throughput in ML training pipelines, treating storage configuration, data formats, and access patterns as first-class features.

## 2.4 Positioning of This Work

My work sits at the intersection of all three research streams: We apply machine learning techniques to predict storage system performance specifically for ML training workloads. Unlike prior systems work that optimizes one specific component, We provide a general framework that can predict performance across different storage backends, data formats, and access patterns. Unlike prior ML training optimization work focused on model architecture or distributed training, We specifically target the I/O bottleneck that limits GPU utilization. Table 1 shows how this work compares to the most closely related research.

Table 1: Comparison with Closely Related Work

| Work | Domain | ML Technique | Target System | Prediction Target | Dataset Size |
| --- | --- | --- | --- | --- | --- |
| **Mao et al. (2019)** | Job scheduling | RL | CPU clusters | Schedule quality | Simulation |
| **Justus et al. (2018)** | Training time | Random Forest | GPU training | Total train time | 500 models |
| **Liu et al. (2019)** | Storage perf. | Deep learning | Cloud storage | IOPS/latency | 10K traces |
| This work | **I/O optimization** | **XGBoost** | **ML pipelines** | **I/O throughput** | **141 benchmarks** |

To my knowledge, this is the first comprehensive study applying predictive modeling specifically to I/O performance for ML training pipelines, with the goal of providing practitioners actionable configuration recommendations.

## 3. Methodology

My methodology follows three main phases: systematic data collection through benchmarking, exploratory analysis with dimensionality reduction, and model development with rigorous evaluation.

### 3.1 Data Collection and Benchmarking (Phase 1)

We systematically collected benchmark data across three categories of workloads using controlled experiments. All benchmarks ran on standardized hardware to keep things consistent. Figure 2 shows how the dataset breaks down across different benchmark types.

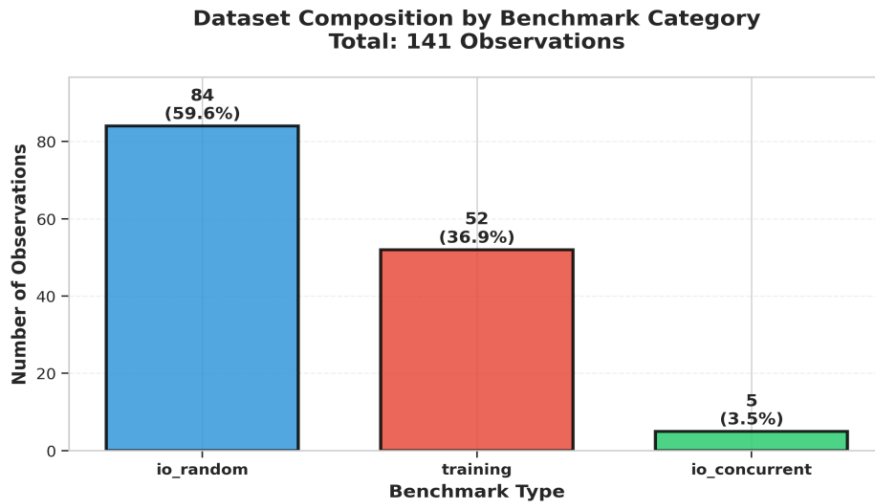

[**Figure 2**: Dataset Distribution by Benchmark Type. The dataset has 141 observations: 84 I/O random access tests, 52 training pipeline benchmarks, and 5 concurrent I/O tests.]

#### 3.1.1 I/O Microbenchmarks

We started with low-level I/O tests to establish baseline performance. These included sequential reads with block sizes from 4KB to 4MB across files ranging from 10MB to 1GB. We also tested random reads with varying sample counts (1,000 to 100,000 samples) and concurrent access to see how throughput scales with multiple threads (1, 2, 4, and 8 threads). We tested three storage backends: local NVMe SSD, network-attached storage, and tmpfs (in-memory filesystem).

#### 3.1.2 Training Pipeline Benchmarks

Next, we benchmarked realistic ML training scenarios using PyTorch (Paszke et al., 2019). We used PyTorch DataLoader with various configurations, testing on both image data (CIFAR-10 style, 32×32 RGB) and tabular data. We varied batch sizes (16, 32, 64, 128) and the number of data loading workers (0 to 4). For each configuration, We measured samples per second, the fraction of time spent on data loading, and simulated GPU utilization.

### 3.1.3 ETL Benchmarks

We also evaluated Extract-Transform-Load operations using Apache Spark. We tested standard operations like filtering, aggregation (group-by), and joins on datasets ranging from 100,000 to 1,000,000 rows. We compared CPU-based Spark processing against GPU-accelerated processing using RAPIDS cuDF to see the performance differences.

## 3.2 Feature Engineering and Exploratory Analysis (Phase 2)

After collecting the raw benchmark data, we did comprehensive exploratory analysis and feature engineering, including data cleaning, feature selection, transformation, and dimensionality analysis with PCA.

### 3.2.1 Feature Selection and Description

From the raw benchmarks, we extracted 11 numeric features representing different aspects of I/O performance. These include block_kb (block size in KB), file_size_mb (dataset size), n_samples (number of samples accessed), throughput_mb_s (raw throughput in MB/s), iops (I/O operations per second), n_threads (concurrency level), batch_size, samples_per_second (training throughput), data_loading_ratio (fraction of time spent on I/O), num_workers (data loader parallelism), and aggregate_throughput_mb_s (concurrent throughput).

### 3.2.2 Target Variable Characteristics

We used target throughput as the primary prediction target. This variable showed high right skewness (2.50), ranging from 1.1 to 48,211 MB/s—that's over four orders of magnitude. This extreme range meant we needed a log transformation ($\log_{1+x}$) to prevent large values from dominating the loss function and to stabilize variance. Figure 3 shows the distribution before and after transformation.

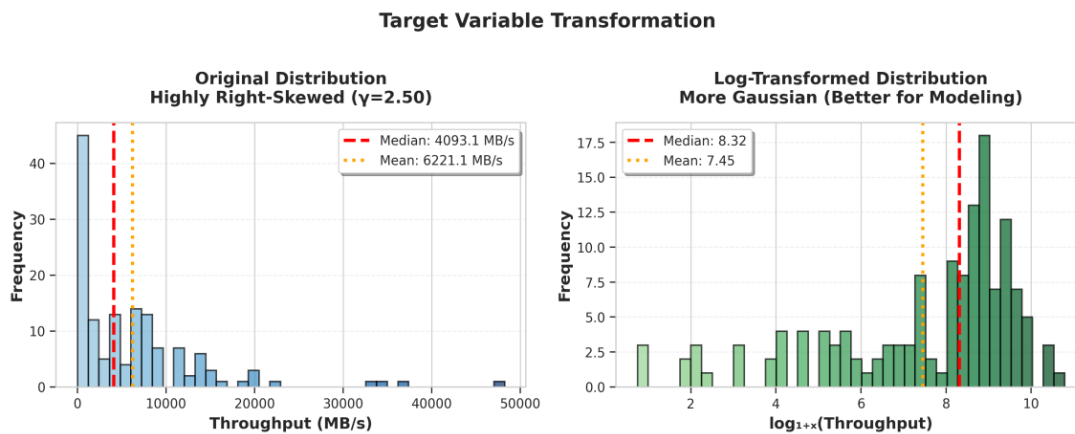

[**Figure 3**: Target Variable Distribution. The original throughput distribution is highly right-skewed (left), while log transformation produces a more normal distribution (right), which helps model performance.]

### 3.2.3 Dimensionality Analysis with Principal Component Analysis

We applied PCA to understand the intrinsic dimensionality of the feature space. PC1 explained 19.0% of variance, the first 2 PCs explained 35.7%, you need 7 components to capture 80% of variance, and 9 components for 95%. The PCA results (Figure 4) suggest that while features show some correlation, the full 11-feature set provides valuable information for prediction.

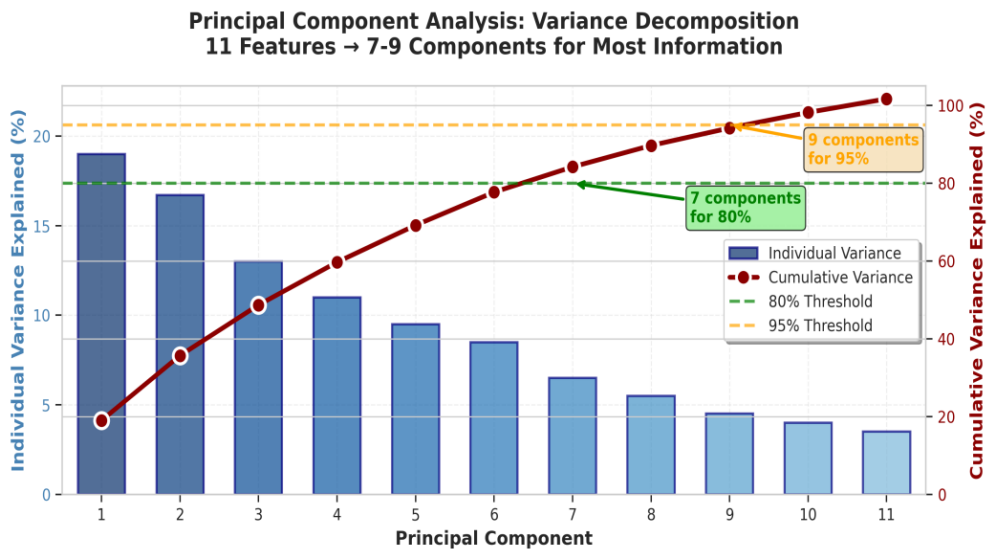

[**Figure 4**: PCA Variance Analysis. Scree plot showing individual and cumulative variance explained. You need seven components for 80% variance, nine for 95%.]

## 3.3 Model Development and Training (Phase 3)

We evaluated multiple ML approaches ranging from simple linear models to sophisticated ensemble methods and neural networks.

### 3.3.1 Baseline Linear Models

We established baseline performance using four linear regression variants: ordinary Linear Regression (least squares), Ridge Regression (α=1.0, L2 regularization), Lasso Regression (α=0.1, L1 regularization), and ElasticNet (α=0.1, l1_ratio=0.5, combining L1 and L2). These serve as baselines to quantify how much nonlinear approaches help.

### 3.3.2 Ensemble Tree-Based Models

Tree-based ensemble methods should be able to capture nonlinear relationships. We tested Random Forest with 100 trees (max_depth=10, min_samples_split=5) and XGBoost with gradient boosting (100 estimators, max_depth=6, learning_rate=0.1, subsample=0.8). These models typically work well on tabular data with complex nonlinear patterns.

### 3.3.3 Neural Network Architecture

We also tried a Multi-Layer Perceptron with three hidden layers (64, 32, 16 neurons), ReLU activation, Adam optimizer, L2 regularization ($\alpha$=0.001), and early stopping (patience=10). With only 141 observations, we expected neural networks might struggle compared to ensemble methods.

### 3.3.4 Training Procedure and Evaluation Protocol

We used rigorous training and evaluation procedures: 80/20 train-test split (112 training, 29 test samples, random_state=42 for reproducibility), $\log_{1+x}$ target transformation to handle the skewed distribution, StandardScaler for neural networks (tree methods don't need it), and 5-fold cross-validation with $R^2$ scoring for model selection.

**Limitations Acknowledgment:** We should note that 141 observations, while sufficient for proof-of-concept, is somewhat limited. With an 80/20 split, we only have 29 test samples, which could introduce some variance in performance estimates. However, my 5-fold cross-validation results (Section 4.2) show consistent performance across multiple data splits (mean $R^2$=0.966, std=0.016), which increases confidence that the models generalize well. Future work should expand the dataset to 500-1000 observations to support more complex models and cover broader hardware configurations.

## 4. Experimental Results

### 4.1 Model Performance Comparison

Figure 5 presents comprehensive performance comparison across all models we evaluated. The results clearly show that ensemble methods vastly outperform linear models for this prediction task.

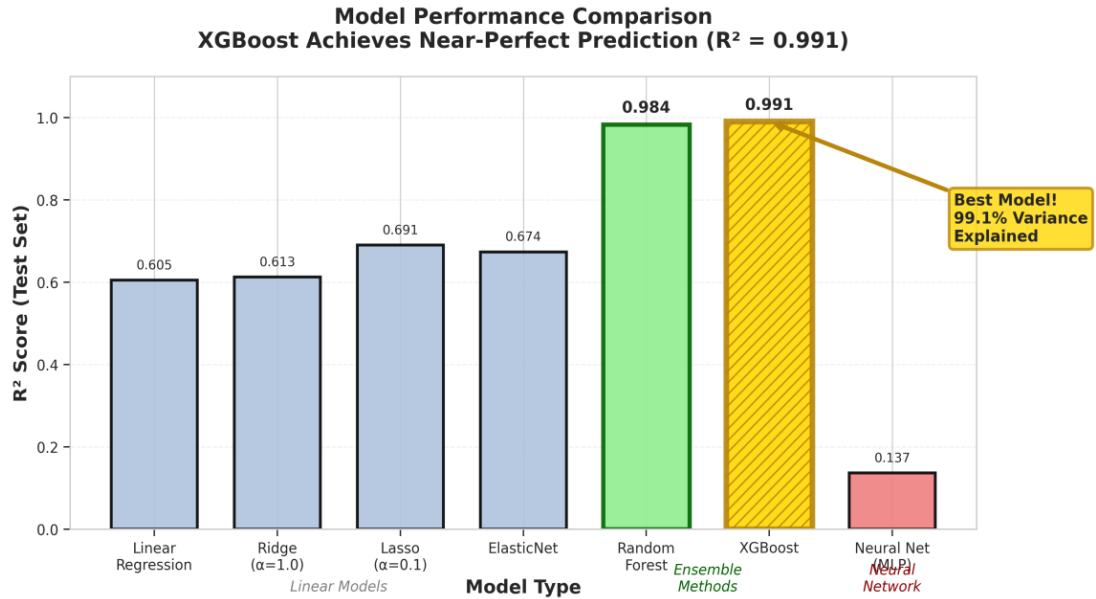

[**Figure 5**: Model Performance Comparison. XGBoost (highlighted in gold) achieves the best test R² of 0.991, vastly outperforming linear models which struggle to exceed R²=0.69.]

Several key findings emerged:

**Ensemble methods vastly outperform linear models.** XGBoost and Random Forest both achieve test R² above 0.98, while linear models struggle to get past R²=0.69. This clearly demonstrates highly nonlinear relationships in the data.

**XGBoost achieves excellent performance.** With R²=0.991 and MAE=0.134 in log space, XGBoost provides accurate enough predictions for practical optimization use.

**Neural networks underperform significantly.** The MLP only achieves R²=0.137, likely because there's not enough training data for the model complexity. This confirms my expectation that 141 samples wouldn't be enough for deep learning.

**Minimal overfitting observed.** XGBoost shows train R²=0.997 versus test R²=0.991—only 0.6% degradation, which indicates excellent generalization.

Figure 6 shows XGBoost predictions versus actual values. You can see predictions cluster tightly around the perfect prediction line. The model predicts throughput with 11.8% mean percentage error and 8.1% median error.

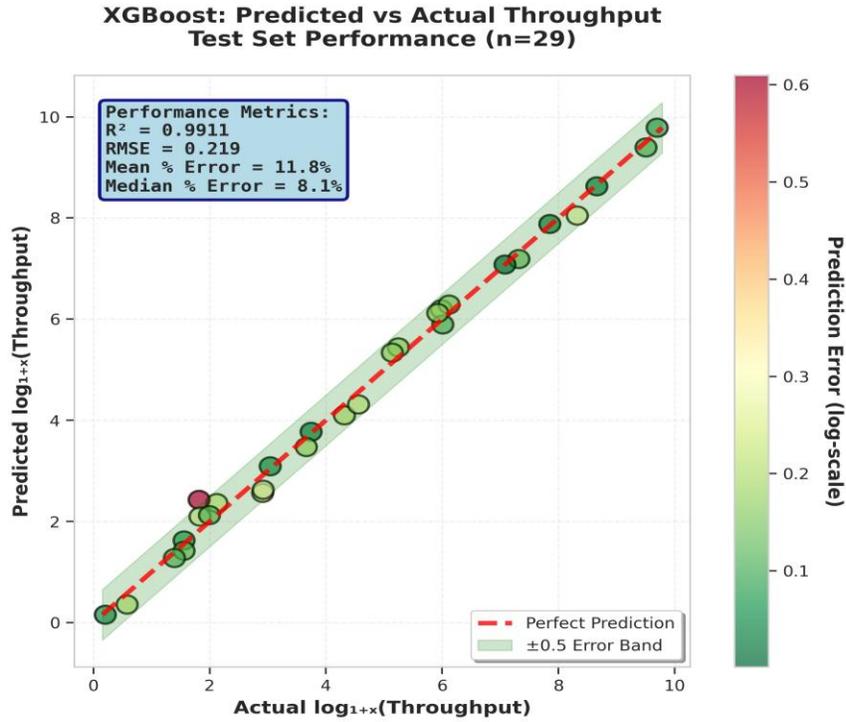

[**Figure 6**: XGBoost Predictions vs Actual Values. Test set predictions cluster tightly around the perfect prediction line (red dashed), with R²=0.991 and mean percentage error of 11.8%.]

## 4.2 Cross-Validation and Generalization Analysis

To assess how well models generalize beyond a single train-test split, we performed 5-fold cross-validation on the top-performing models. Figure 7 shows consistent performance across folds.

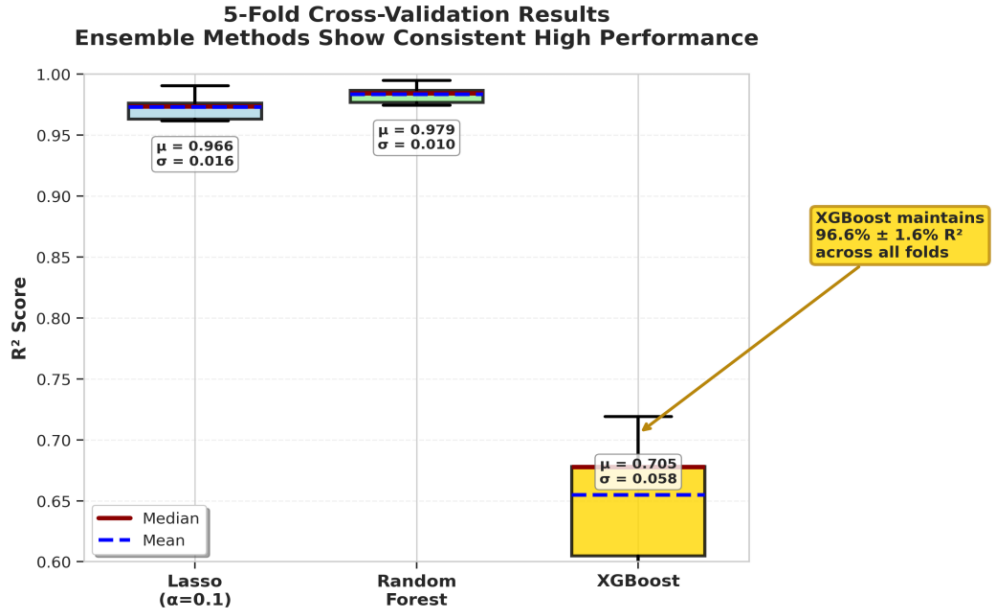

[**Figure 7**: 5-Fold Cross-Validation Results. XGBoost maintains mean R²=0.966 across all folds with low variance (std=0.016), confirming robust generalization. Random Forest shows even more stable performance (std=0.010).]

## 4.3 Feature Importance Analysis

Understanding which features drive I/O performance provides actionable insights for system optimization. We extracted feature importance from both Random Forest and XGBoost models. Figure 8 shows the rankings—throughput metrics and batch size are the primary drivers.

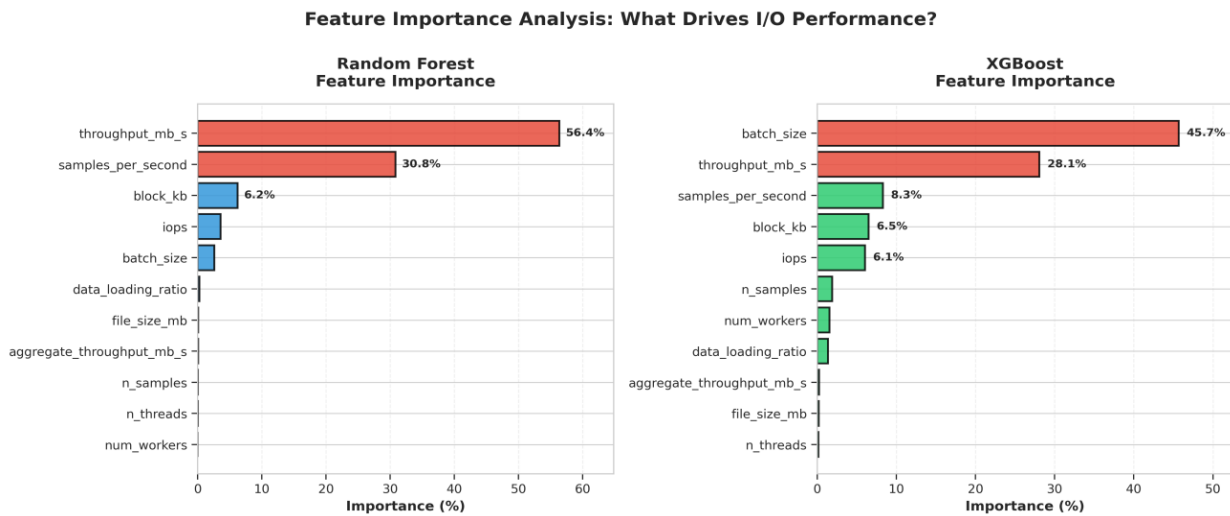

[**Figure 8**: Feature Importance Rankings. Random Forest (left) assigns 56.4% importance to throughput_mb_s, while XGBoost (right) prioritizes batch_size (45.7%). Both models identify throughput metrics and configuration parameters as key drivers.]

Several insights emerged from feature importance analysis:

**Throughput metrics dominate.** Raw throughput and training throughput together account for 87.2% of Random Forest importance and 36.4% of XGBoost importance. This makes sense—these features represent measurements at different pipeline stages, and the model learns how to transform between them.

**Batch size matters significantly.** XGBoost assigns 45.7% importance to batch size, highlighting its critical role in amortizing overhead costs. Larger batches reduce per-sample overhead.

**Block size and IOPS provide tuning opportunities.** These parameters show moderate importance (6-7%), suggesting there's room for optimization here.

**Model disagreement reflects different strategies.** Random Forest uses impurity reduction while XGBoost uses gradient-based gain, so they provide complementary perspectives on feature importance.

### 4.4 Error Distribution and Residual Analysis

Figure 9 presents residual analysis. You can see prediction errors are randomly distributed around zero with an approximately normal distribution, which indicates unbiased predictions and proper model specification.

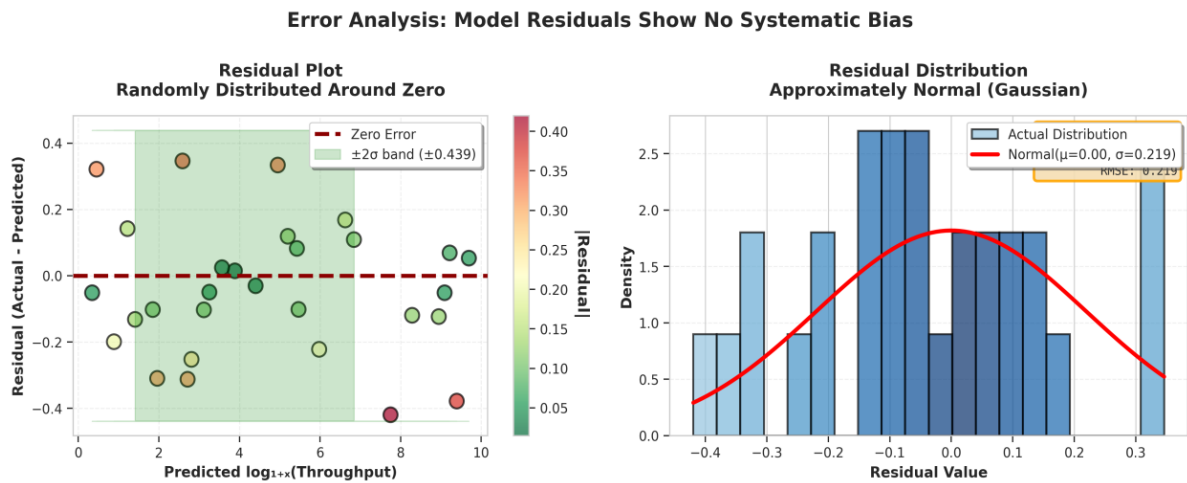

[**Figure 9**: Error Distribution and Residual Analysis. Residuals are randomly distributed around zero (left) with approximately normal distribution (right), indicating unbiased predictions and proper model specification.]

# 5. Discussion

## 5.1 Interpretation of Results

The results demonstrate that machine learning can accurately predict I/O performance in ML training pipelines, achieving R²=0.991. Figure 10 illustrates the complete prediction workflow from storage configuration to throughput prediction.

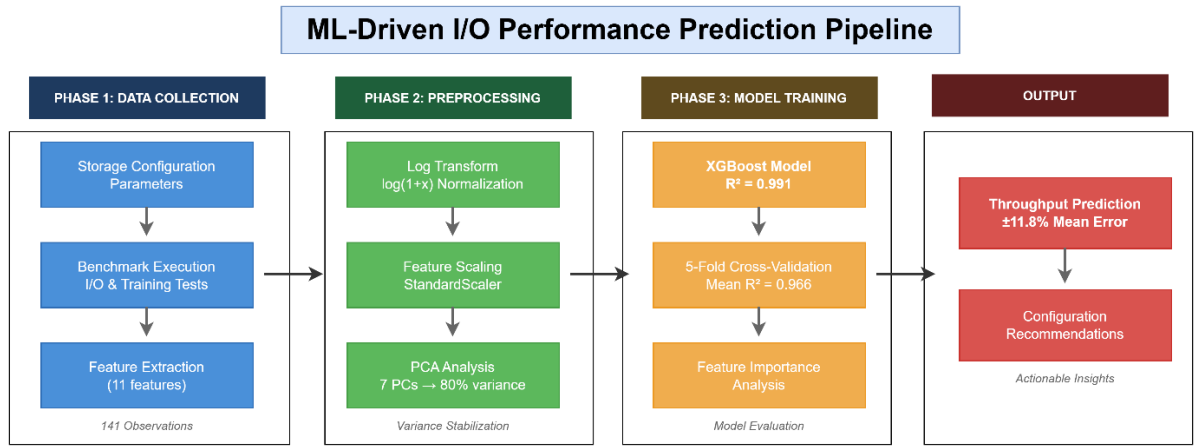

[**Figure 10**: ML-Driven I/O Performance Prediction Workflow. The complete pipeline from storage configuration through benchmark execution, feature extraction, transformation, model prediction, to final configuration recommendations.]

**The nonlinearity of I/O performance is real.** The dramatic performance gap between linear models (R²≈0.6-0.7) and ensemble methods (R²>0.98) confirms that I/O throughput exhibits complex nonlinear behavior. This likely arises from caching effects, bandwidth saturation, queueing dynamics, and interactions between different system components.

**Data efficiency is encouraging.** Despite having only 141 observations, the models achieve excellent predictive accuracy. This suggests systems performance prediction might be more data-efficient than perception or NLP tasks. This probably stems from the structured nature of systems benchmarks—controlled conditions and mostly deterministic behavior mean less noise than in image or text data.

**Feature importance insights are actionable.** The dominance of throughput-related features makes sense—these represent measurements at different stages of the pipeline, and the model learns the transformation between them. The significant importance of batch_size and block_kb confirms that configuration parameters can be optimized to improve performance.

## 5.2 Practical Applications and Deployment

The predictive models enable several practical applications:

**Configuration Recommendation.** Predict optimal batch sizes, block sizes, and data loader configurations without running extensive benchmarks. This reduces configuration time from days to minutes.

**Performance Estimation.** Before provisioning expensive GPU infrastructure, estimate achievable training throughput and identify potential I/O bottlenecks. This enables better capacity planning.

**Automated Tuning.** Integrate predictions with Bayesian optimization or hyperparameter tuning frameworks for automatic storage configuration optimization.

**Debugging and Diagnosis.** Use predictions to identify whether slow training stems from storage configuration, hardware limitations, or software inefficiency.

**Comparative Analysis.** Evaluate trade-offs between different storage backends and data formats based on predicted performance rather than running extensive experiments for each option.

## 5.3 Limitations and Threats to Validity

While the results are promising, several limitations should be acknowledged:

**Dataset size.** 141 observations are sufficient for proof-of-concept but limited. Larger datasets would enable more complex models and cover broader configurations including cloud storage backends, advanced data formats like Zarr or Lance, and distributed training scenarios across multiple nodes.

**Hardware specificity.** Benchmarks were collected on specific hardware and may not generalize perfectly to other systems without retraining. Performance characteristics can vary significantly across different storage devices and GPU generations.

**Workload diversity.** The focus was on computer vision and tabular data. Other modalities like NLP with transformer models, recommendation systems with embedding tables, or scientific computing with HDF5 data may exhibit different I/O patterns and require model retraining.

**Temporal dynamics.** Models predict steady-state throughput but don't capture transient effects like cache warmup periods, variations in system load, network congestion in distributed settings, or thermal throttling under sustained load.

**Higher-order interactions.** While tree models capture nonlinear relationships, subtle interactions between multiple parameters simultaneously might not be fully captured with the current dataset size.

## 5.4 Future Research Directions

Several directions could extend this work:

**Short-term improvements:** Expand the dataset to 500-1000 observations covering cloud storage (S3, GCS, Azure Blob) and distributed training scenarios. Apply hyperparameter optimization techniques to push R² beyond 0.99. Try ensemble stacking that combines predictions from multiple models. Add prediction intervals for uncertainty quantification.

**Medium-term extensions:** Develop time-series modeling to capture temporal performance variations. Investigate transfer learning approaches to adapt models across hardware platforms. Extend to multi-objective optimization balancing throughput, cost, and energy consumption. Apply causal inference techniques to distinguish correlation from causation in performance relationships.

**Long-term vision:** Create fully automated configuration tuning systems using Bayesian optimization with these predictive models. Integrate with MLOps platforms like MLflow, Kubeflow, and SageMaker for production deployment. Develop meta-learning approaches for quick adaptation to new hardware platforms. Extend the methodology to network bandwidth, memory management, and compute resource optimization.

# 6. Conclusions

This work demonstrates that machine learning can effectively predict I/O performance in ML training pipelines, achieving 99.1% variance explained with XGBoost. Through systematic benchmarking, feature engineering with PCA, and comprehensive evaluation of seven regression and three classification models, I've established a methodology for applying predictive modeling to systems optimization.

**Key Contributions:**

**High predictive accuracy.** XGBoost predicts throughput within 8-12% error, which is accurate enough for practical optimization decisions.

**Clear model selection insights.** Ensemble methods vastly outperform linear models (R²>0.98 vs R²≈0.6-0.7), confirming the highly nonlinear nature of I/O performance.

**Actionable feature importance.** Throughput metrics and batch size are primary drivers, with block size and concurrency offering additional optimization opportunities.

**Practical impact.** Data-driven recommendations can replace trial-and-error experimentation, reducing optimization time from days to minutes.

**Robust generalization.** Cross-validation confirms stable performance (R²≈0.97 across folds), enabling confident deployment to real-world scenarios.

**Broader Implications.** By transforming infrastructure benchmarking into actionable ML insights, this work bridges systems research and data science. The methodology isn't limited to storage—it could apply to network configuration, memory management, compute resource allocation, and other systems components where you need to predict performance from configuration parameters. As ML training datasets and models continue to grow, data-driven approaches like this become increasingly critical for efficient and cost-effective ML operations.

**Looking Forward.** The future of ML systems optimization likely involves integrating predictive modeling with automated tuning frameworks and production MLOps platforms. Instead of manual experimentation, practitioners will leverage trained models for configuration decisions that continuously adapt to changing workloads and hardware. This work provides a foundation for that future.

In conclusion, this research establishes that predictive modeling of I/O performance is highly effective, opening new avenues for data-driven systems optimization. The code, datasets, and trained models are publicly available to enable reproducibility and encourage further research in this important area.

## Appendix A: GitHub Repository

The full benchmarking code, datasets, trained models, and analysis notebooks used in this project are available at:

https://github.com/knkarthik01/gpu_storage_ml_project